\documentclass[pre,twocolumn,superscriptaddress]{revtex4-1}

\usepackage{subfigure}
\usepackage{graphicx}
\usepackage{epsfig}
\usepackage{amsmath}
\usepackage{epstopdf} 
\usepackage{color}
\usepackage{mathtools}
\usepackage{float}

\usepackage{soul}

\usepackage{hyphenat}
\usepackage[colorlinks=true,linkcolor=blue,urlcolor=blue,citecolor=blue]{hyperref}

\newcommand{\bracketed}[1]{\left[\vphantom{\let\\=\relax #1}\right. #1 \left.\vphantom{\let\\=\relax #1}\right]}

\begin{document}
\title{DNA melting in poor solvent}

\author{Debjyoti Majumdar}
\email{debjyoti@iopb.res.in}
\affiliation{Institute of Physics, Bhubaneswar, Odisha 751005, India}
\affiliation{Homi Bhabha National Institute, Training School Complex, Anushakti Nagar, Mumbai 400094, India}

\date{\today}

\begin{abstract}
The melting phase diagram of a double-stranded DNA in poor solvent is studied using the pruned and enriched Rosenbluth method on a simple cubic lattice. As the solvent quality is changed from good to poor, there is a non-monotonic change in the melting temperature. First-order melting transition, as in good solvent, gives way to continuous transition and then to further broadened transitions where the order parameter smoothly becomes zero for sufficiently poor solvent. This change in the melting behavior is accompanied by a continuously varying critical exponent along the melting curve, hinting at a non-universal nature of the melting transition. Further, we show that an unbound phase can be achieved just by changing the solvent quality. Importantly, our results conform to the experimental findings qualitatively.
\end{abstract}

\maketitle

\section{Introduction} 
The double-stranded (ds) DNA consists of two polymer strands connected along the contour {by hydrogen bonds between the {bases forming base-pairs (bps)} \cite{watson1953}. {The genetic information is encoded in the sequence of these bases. {Biological processes like gene expression and replication require the separation of these two strands and then copying the genome sequence}. {\it In vitro}, this separation of the dsDNA into two single-strands (ss) can be carried out by changing the temperature and is known as the {\it thermal melting}} {transition}. {In theoretical studies}, the melting transition is generally considered in the context of a linear structure of DNA, {a famous example being the} Poland-Scheraga (PS) model \cite{poland1966}. The strands, being polymers by {themselves}, can further undergo a significant reduction in size through the {\it condensation} transition, either in the ds (bound) or ss (unbound) phase, when exposed to sufficiently poor solvent. {It is this condensed form, in which,} DNA is generally found {\it in vivo}, e.g., in prokaryotes  \cite{comm1}. 

Condensation leads to the formation of higher-order structures through loops and domains \cite{joyeux2015}. These higher-order structures are instrumental in modifying the melting transition due to a change in the accessibility among the inter- and intra-strand segments of the DNA \cite{bouligand1985,sikorav1991}. Further, condensation promotes long-range order $\mathcal{O}$(cellular dimensions), essential for {biological processes like} morphogenesis {and also with a possible role in optimizing the renaturation rate} \cite{sikorav1991}. Moreover, since changing the temperature {\it in vivo} is not a feasible option, it is the solvent that has  to play a key role in controlling the state of the DNA, e.g., for enzymes, changing the local micro-environment of the DNA could be a mechanism for altering the bps stability locally \cite{cui2007}. Therefore, it is instructive to study the DNA melting transition in light of variable solvent quality.

For DNA, the polarity of the solvent determines the solvent quality, with all good solvents having high polarities, e.g., water, since DNA is also a highly polar object, {owing to the negatively charged backbone}. {Several experimental studies have been performed in the last four decades \cite{baldini1985,hammouda2009,feng1999,potaman1980, rupprecht1983, mikhailenko2000, cui2007}, studying the role of solvents in DNA melting. A common observation is that} an increase in the alcohol concentration {in the DNA solution \cite{baldini1985}} {or increasing the carbon content in the hydrophobic tail of the alcohol} \cite{hammouda2009}, tends to decrease and then increase the melting temperature. While the decrease in the melting temperature  is poorly understood, the increase in the thermal stability can be attributed to the aggregation and precipitation of the bound DNA in the presence of alcohol \cite{potaman1980, rupprecht1983,feng1999}. {It was shown that} melting occurs mainly in the region where DNA has an elongated conformation, and therefore, there exists a significant cooperative effect between the melting and the unfolding transition \cite{mikhailenko2000}. {In Ref. \cite{cui2007}, Cui {\it et al.} found that at room temperature, dsDNA can be denatured into ssDNA in organic poor solvents using AFM-based single molecule force measurement experiments. This observation was further supported by molecular dynamics simulations, where the double-stranded form split spontaneously into two single strands} when pulled at one strand from an aqueous to a poor solvent region. {On the theoretical side, efforts were made for DNA on special lattices to incorporate the solvent effect}, e.g., in Ref.~\cite{foster2009}, the zipping/unzipping transition of DNA in variable solvent quality is studied using the Bethe lattice. {Again, using a non-linear model of the DNA,} known as the Dauxois–Peyrard–Bishop model, it was shown that the melting temperature decreases as the solvent potential increases \cite{macedo2014}.

{Although several experimental papers exist investigating the effect of solvent quality on DNA melting, it seems that a precise reason explaining why the melting curve behaves the way it was found in Ref.~\cite{baldini1985} is lacking. Experimental difficulties in quantifying the interactions among segments from different phases may have hindered the development of a quantitative understanding. This is where the role of computer simulations is important.} To the best of our knowledge, a proper account of the solvent quality, along with the base-pairing effect in the proper dimension, using computer simulation of lattice models, is missing in the literature. Therefore, we believe that further work is needed for a {complete} understanding.

In this article, we propose a model for the condensation of a melting DNA and show that for a flexible DNA, of finite length, with complementary base-pairing, the {temperature induced} melting transition is completely modified in a poor solvent background \cite{comm7}. Interestingly, our simple coarse-grained lattice model, which includes solvent only implicitly, through an effective attraction among different segments {(bound and unbound)}, is capable of capturing qualitatively the features found previously in the experiments \cite{baldini1985, mikhailenko2000}. {Further, we address several aspects absent in the previous studies,} {such as bubble statistics}. The finiteness of the DNA in our model makes it relevant from an experimental standpoint, since experiments are performed upon finite systems.

This paper is organized in the following manner: in Sec.~\ref{sec2}, we describe the DNA model for simultaneously studying the melting and condensation phenomena. {In Sec.~\ref{sec3}, we describe the method for simulating the {dsDNA} on the cubic lattice}. In Sec.~\ref{sec4}, we discuss the findings, and conclude the paper in Sec.~\ref{sec5}.

\section{DNA model and qualitative description}  
\label{sec2}
In our study, {we use a coarse-grained model} of the DNA \cite{causo2000}, {where the two strands of the dsDNA (say strands A and B) {are} represented by two self- and mutually-avoiding linear polymers}, embedded on a cubic lattice in the dilute limit. One of the ends of the DNA is fixed at the origin, while the other end remains free to wander [Fig.~\ref{fig1}(a)]. {Each strand contains $N$ monomers, with each monomer acting as a base and capable of forming a base-pair (bp) of energy $-E_{bp}$, with the complementary monomer (same monomer index) along the other strand, while both the monomers occupy the same lattice site}. {Apart from the bp interaction}, there is an effective attraction due to the poor solvent background, between the non-bonded inter and intra-strand segments, with an energy  $-E_{nn}$ when one lattice constant apart \cite{grassberger1997}; a condition referred to as the nearest-neighbor (nn). In configurations where a contact site of the two strands (i.e., a ds segment) is nn with another occupied lattice site, which is part of either ss or ds, {we consider} only one nn interaction [Fig.~\ref{fig1}(b)]. Such a consideration arises from the fact that any difference in the collapse of the ds and ss segments should arise naturally from the difference in the rigidity of them, the former being $25$ times {more} rigid than the latter, in reality.

\begin{figure}[t]
\centering
\hspace*{-7cm}(a)\\
\includegraphics[width=0.7\linewidth]{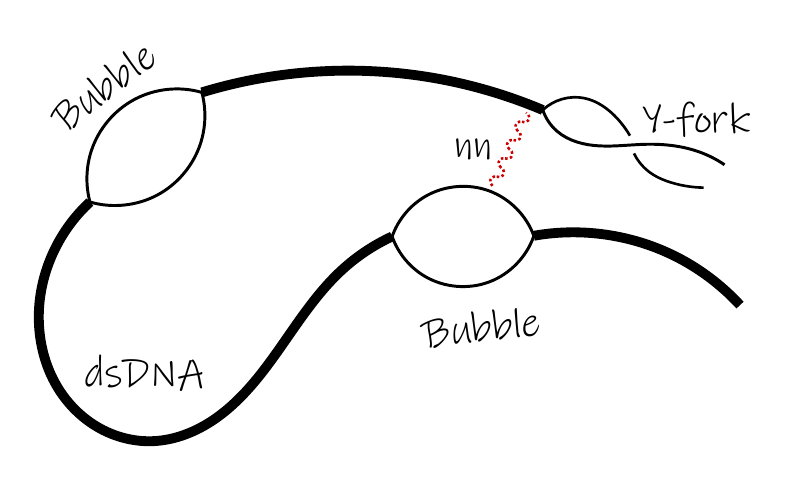}\\
\hspace*{-7cm}(b)\\
\includegraphics[width=0.7\linewidth]{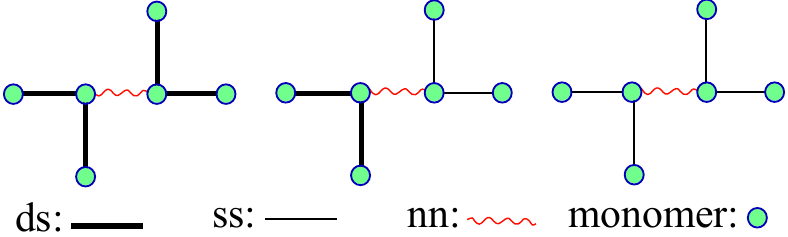}
\caption{(Color online) (a) Schematic representation of our DNA model with nn representing nearest-neighbor interaction due to the poor solvent (red squiggly lines). Broken bonds enclosed within bound segments are referred to as {\it bubbles}, and as Y-fork if at the end. (b) {Possible two-step  configurations contributing to non-bonded nn and the associated energies. The total bp energy is $-6$, $-3$ and $0$ in {the} unit of $E_{bp}$, for the three figures (left to right), respectively. The total nn interaction energy is $-1$ in {the} unit of $E_{nn}$ for all three cases.}}
\label{fig1}
\end{figure}

The model described above contains all the necessary details required to study the melting and collapse transition individually and also the interplay of the two. However, for simplicity, we have neglected the rigidity of the ds segments, along with certain other aspects of the dsDNA, such as its helical structure and sequence heterogeneity, under the assumption that the effect of these {properties} can be studied separately. Moreover, {considering} all of the factors in a single simulational study would substantially limit the maximal system size.

{Our model DNA can be decomposed into} three parts: {bound segments}, broken bonds enclosed within bound segments known as {\it thermal bubbles}, and a {Y-fork} at the free end [Fig.~\ref{fig1}(a)]. {This categorization of the DNA into three parts is important because, as we will see, the interplay of the nn interactions among these three segments will determine the behavior of the melting curve}. A schematic representation of our model is shown in Fig.~\ref{fig1}(a) and {the} scheme for assigning energy to the non-bonded nn in Fig.~\ref{fig1}(b). For a typical configuration [Fig.~\ref{fig1}(a)], the Hamiltonian can be written as:

{
\begin{multline} \label{eqn1}
\mathcal{H} = -E_{bp} \sum_{i=1}^{N} \delta_{{\bf r}_i^{\rm A},{\bf r}_i^{\rm B}}   
- E_{nn} \Bigl\lbrace \sum_{i=1}^{N-2}\sum_{j=i+2}^N \bigl( (\delta_{1,\mid {\bf r}_i^{\rm A} - {\bf r}_j^{\rm A} \mid}  \\ + \delta_{1,\mid {\bf r}_i^{\rm B} - {\bf r}_j^{\rm B} \mid} )  \times (1-\delta_{{\bf r}_i^{\rm A},{\bf r}_i^{\rm B}} \delta_{{\bf r}_j^{\rm A},{\bf r}_j^{\rm B}}) + \delta_{1,\mid {\bf r}_i^{\rm AB} - {\bf r}_j^{\rm AB} \mid} \\ \times \delta_{{\bf r}_i^{\rm A},{\bf r}_i^{\rm B}} \delta_{{\bf r}_j^{\rm A},{\bf r}_j^{\rm B}} \bigr) + \sum_{i,j=1}^N \delta_{1,\mid {\bf r}_i^{\rm A} - {\bf r}_j^{\rm B} \mid} \times (1-\delta_{{\bf r}_i^{\rm A},{\bf r}_i^{\rm B}}) (1-\delta_{{\bf r}_j^{\rm A},{\bf r}_j^{\rm B}}) \Bigr\rbrace,
\end{multline} 
}

{where ${\bf r}_i^{\rm A}$ is the position vector of the $i$th monomer of strand A, AB denote ds sites which are parts of both A and B, and $\delta_{ij}$ is the Kronecker delta. Notice that each term in Eq.~\ref{eqn1} is symmetric under exchange of strand tag.}

\begin{figure}[t]
\includegraphics[width=.7\linewidth]{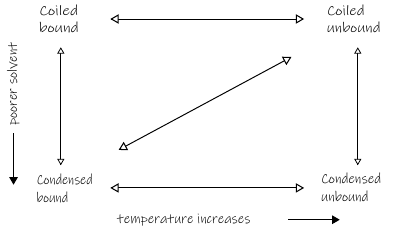}
\caption{Different possible phases and transition pathways. {\it Is a diagonal line possible?}}
\label{fig2}
\end{figure}

The lattice polymer model, which we use to model the DNA strands, can also be used as a model for polymer collapse in the poor solvent with appropriate interactions, known as the interacting self-avoiding walk (ISAW) \cite{grassberger1997}. The collapse of a polymer in three dimensions is a {continuous phase transition} \cite{grassberger1997}, with mean-fieldlike critical exponents at the $\theta$-point, e.g., the exponent $\nu=1/2$ (except for logarithmic corrections), as for the {ideal} random walk, since the excluded volume interaction is exactly canceled out by the nn interactions with the vanishing of the second virial coefficient {for infinitely long chains}. {In the dilute limit,} a single polymer chain in three dimensions collapses at $T=3.717$ for $E_{nn}=1$, or equivalently at $E_{nn}/T=0.269$. To look at the collapse transition, one {can observe} the fluctuation of the nn interactions. In our case this nn interaction  can be divided into three parts which contributes differently in {modifying} the melting transition; among ss segments ($nn_{\text{ss}}$), among ss and ds segments ($nn_{\text{ss-ds}}$), and among ds segments ($nn_{\text{ds}}$) only.

When we consider base-pairing and condensation together, naively, one would expect four different types of phases : (i) condensed-bound phase, (ii) condensed-unbound phase, (iii) coiled-unbound phase, and (iv) coiled-bound phase [Fig.~\ref{fig2}]. Generally, we characterize the bound (or unbound) phase by the average number of bound bps per unit length ($n_c$) being $1(\text{or}~0)$. The condensed phase is characterized by the scaling of the globule size ($R$) with the contour length ($N$), i.e., $R\sim N^{1/d}$ where $d$ is the dimension of the embedding lattice, and the usual SAW exponent in the coiled phase $R\sim N^{\nu}$ with $\nu=0.588$ in three dimensions. Each of these phases, as shown in Fig.~\ref{fig2}, can be reached from the other by moving along any axis. What is interesting {\it if there is any pathway connecting phases (i) {to} (iii) ?}, which corresponds to the diagonal line in Fig.~\ref{fig2}. This will indicate a transition from a bound-collapsed to an unbound-coiled state. {In such a scenario, the major difference in melting will be due to the emergence of the globule surface tension. Such a transition might be important from {the} biological point of view; however, it is out of the purview of the present model. Therefore, we mention it only as a possibility and leave it for future studies. Note that, for a flexible DNA, the completely bound and unbound phases are equivalent, except for a length factor of 2, when considered {concerning} a change in the nn interaction energy $E_{nn}$ at a fixed temperature $T$}.

\section{Simulation algorithm} 
 \label{sec3}
 
For simulation, we use {a sequential Monte Carlo algorithm, called the} pruned and enriched Rosenbluth method (\textsc{PERM}) \cite{grassberger1997, causo2000}. The chains are built in steps, with a recursive call to the \textsc{perm()} subroutine, adding a monomer on top of the lastly added monomer, both the strands at once. This is possible since we know the free sites for the next step of both the strands and a null atmosphere \cite{comm3} would result in an immediate exit of the current call to the subroutine. The joint possibilities arising from the individual atmospheres is $atmos=atmos_{\rm A}\times atmos_{\rm B}$, since, at any point of growth, the individual strands are of the same length. At $n$th step, we calculate the following quantity 
\begin{equation}
w_n=\sum_{atmos} p^k \times q^l,
\end{equation}
where $p=\exp(\beta E_{bp})$ and $q=\exp(\beta E_{nn})$  are the Boltzmann factors for {the} bp and {the} nn interaction respectively, and $k$ and $l$ are their respective numbers. While $k$ take values of $0$ or $1$, $l$ can take values ranging from $0,1\cdots (2d-1)$, where $d(=3)$ is the dimension of the underlying lattice. The final choice is made among $atmos$ number of choices in accordance with the {\it importance sampling} scheme. The total weight of a chain of length $N$ is $W_N(\beta)=\prod_{n=1}^{N} w_n$, from which we obtain the estimate of the partition sum $Z_N(\beta)=\langle W_N(\beta) \rangle$, the average being taken over the number of started tours \cite{comm4}. Pruning and enrichment {at $n$th step} is performed depending upon the ratio, $r=W_n/Z_n$. Different schemes can be employed in choosing the enrichment threshold and the number of copies. However, we find the algorithm to be robust in its performance. {In our simulations we have used; $r>1$ as the enrichment threshold, with $c$ different copies each with weight $\frac{W_n}{c}$ and $c=min(\lfloor r \rfloor,atmos)$, present configuration continues to grow for $r=1$, and is pruned with probability $1-r$ if $r<1$ or continue to grow with probability $r$ but with $W_n=Z_n$} \cite{prellberg2004}. We take averages over $10^7$ tours {and} have set $E_{bp}=k_B=1$ {in all of our simulations.}

For monitoring lattice site occupancy, we use the method of {\it bit maps}, where the whole cubic lattice is stored into a one-dimensional array, and, {occupied lattice sites are indexed 1(or 2) if occupied by strand A (or B), 3 if occupied by ds, and 0 if empty}. Thus, it facilitates keeping a count of the ss-ss, ss-ds, and ds-ds nn interactions separately, with an added advantage that the time complexity for self-avoidance check reduces significantly compared to any other method, e.g., binary search tree. {However,} this added advantage comes at the cost of a finite lattice box size, and one has to be careful about the box boundaries. However, in our simulations, {we do not expect extended configurations, e.g., if an external force pulls the strands along a direction,} {in which case {\it linked list} methods will be a better choice.}

\begin{figure}[t]
\includegraphics[width=.9\linewidth]{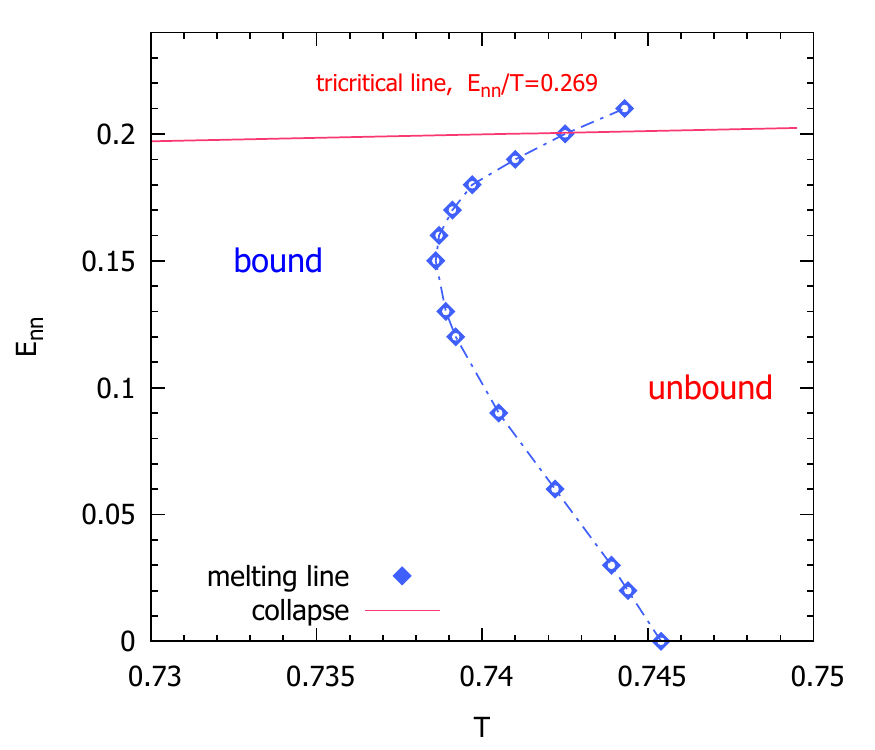}
\caption{(Color online) Phase diagram elucidating the interplay of melting and condensation transition. {Temperature ($T$) is varied along the horizontal axis and solvent quality ($E_{nn}$) is varied along the vertical axis, with $E_{bp}=1$ fixed.} The blue dashed line is a guide to the eye, interpolating the melting points ($T_m$). {To the right, we have the high-temperature unbound phase, and the low-temperature bound phase on the left of the melting curve. Error bars in $T_m$ are of the size of the data points. The numerical determination of the melting points is discussed in Sec.~\ref{sec4}.} The diagonal continuous red line represents the tricritical collapse for an interacting self-avoiding walk (ISAW) which satisfies $E_{nn}/T=0.269$ at all points \cite{grassberger1997}.}
\label{fig3}
\end{figure}
\section{Results and discussion} 
\label{sec4}
In good solvent (i.e. $E_{nn}=0$), our model exhibits a first-order melting transition at {$T_m=0.7455$} for $E_{bp}=1$ \cite{causo2000}. The $\theta$ temperature for the collapse transition {of a lattice polymer in three dimensions} is {$T_{\theta}=3.717$} for $E_{nn}=1$ \cite{grassberger1997}, in unit of Boltzmann constant ($k_B=1$) for both melting and collapse. {Our simulations reproduced these individual results by setting either $E_{nn}=0$ or $E_{bp}=0$, respectively.} In Fig.~\ref{fig3}, we plot the phase diagram in the $E_{nn}$ vs. $T$ plane, with $E_{bp}=1$ fixed. {Horizontally, we plot the melting transition induced by temperature ($T$), and vertically, the collapse transition brought about by changing the {strength of} the nn interaction ($E_{nn}$)}. We estimate the melting point temperature $(T_m)$ and the associated critical exponent ($\phi$) from the scaling plots of the order-parameter $(n_c)$, and its fluctuation ($C_c$) [Fig.~\ref{fig4}]. {The error bars in $T_m$ are obtained from the sensitivity of the scaling plots to the $T_m$ value.} The average binding energy per unit length $(N)$ {should scale} as $n_c\sim N^{\phi-1}g[(T-T_m)N^\phi]$, while its fluctuation {should scale as $C_{c}\sim N^{2\phi-1}h[(T-T_m)N^\phi]$, where $g$ and $h$ are scaling functions. Tuning $T_m$ and $\phi$ to the appropriate values would then make the data $N$ independent, resulting in data-collapse, as in} Fig.~\ref{fig4}. The crossover exponent $\phi$ controls the sharpness of the {melting} transition, with higher values representing sharper transitions. {For $\phi = 1$, the transition is categorized as first-order and continuous if $\phi<1$}. {In the following paragraphs, we present our main findings, and then try to explain them}.

As we increase the nn attraction ($E_{nn}$), initially, the melting temperature ($T_m$) decreases linearly with the solvent parameter $E_{nn}$ [Fig.~\ref{fig3}]. Then, after a point $(E_{nn} \approx 0.15)$, $T_m$ starts increasing, and the melting curve changes slope as it approaches the tricritical collapse line. This {behavior} is qualitatively similar to what was observed in the experiments \cite{baldini1985, mikhailenko2000, hammouda2006}. Baldini et al. \cite{baldini1985} have further found a change in the overall nature of the melting curve with a different solvent type. However, our model allows change only up to an effective attraction between monomers, be it for another solvent or a different concentration of the same solvent. The change in the melting temperature ($T_m$) is further associated with a {continuous} decrease in the value of the exponent $\phi$ as $E_{nn}$ increases [Fig.~\ref{fig4}(a) inset]. {This indicates a change from first-order to continuous melting transition. In Fig.~\ref{fig4}(b) inset, we plot the variation of the peaks of $C_c$ curves with length $N$, for $E_{nn}=0.06,0.13$ and $0.17$. $\phi$ extracted from the slope of these curves are consistent with that of the scaling plots (not shown). This further corroborates a change in $\phi$. } 

The $\theta$ point for the collapse transition is obtained by fixing $T$,  and varying $E_{nn}$. We use three different $T$'s with respect to the good solvent case; in the bound phase $(T=0.71)$, near the melting point $(T=0.741)$, and in the unbound phase $(T=0.75)$. We found that the $\theta$ point is merely affected by the presence of the base pairing interactions. This is evident from a comparison between the nn fluctuation in our DNA model, with that of a single SAW of the same {linear} length in three dimensions, and interacting with the same ratio of $Enn/T$. Therefore, we draw a diagonal-straight line satisfying $E_{nn}/T=0.269$ for the collapse transition in the phase diagram [Fig.~\ref{fig3}]. The melting temperature corresponding to $E_{nn}=0.2$ is $T_m=0.742(5)$ with $\phi=0.70(8)$. Therefore, $E_{nn}/T_m=0.269$ which is {just} over the tricritical collapse line, {is the meeting point of the melting and collapse curves}. 
\begin{figure}
\centering
\hspace*{-7cm}(a) \\
\includegraphics[width=.9\linewidth]{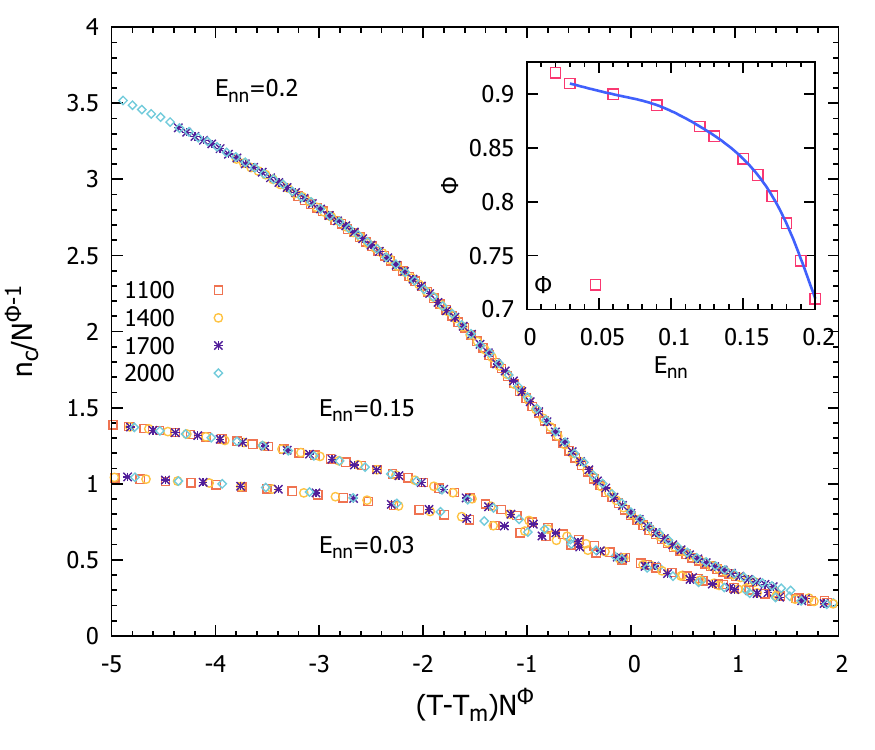}\\
\hspace*{-7cm}(b)\\
\includegraphics[width=.9\linewidth]{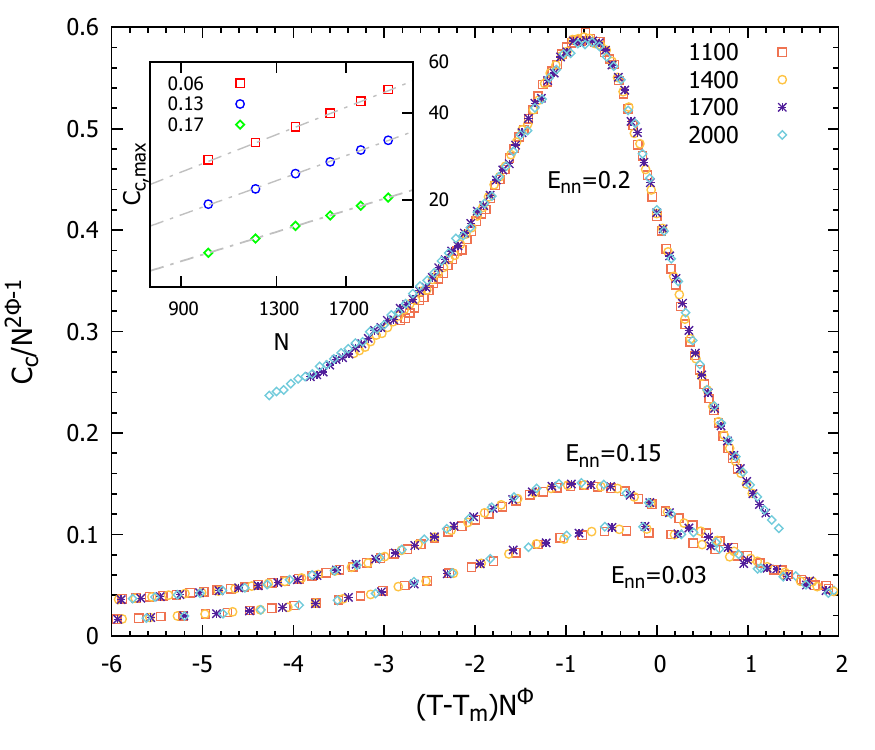}
\caption{(Color online) Scaling plots for (a) average number of bound bps per unit length ($n_c$), and (b) its fluctuation ($C_c$). For (a) and (b), we use $\phi=0.91, 0.84$ and $0.70(8)$, and $T_m = 0.743(9), 0.738(9)$ and $0.742(5)$ for different values of the nearest neighbor energy $E_{nn}=0.03,0.15$ and $0.2$, respectively. Data shown is for chain lengths $N=1100$ to $2000$ at an interval of $300$. (a) Inset: Crossover exponent $\phi$ used in the scaling plots of the melting transition for different $E_{nn}$ values. The continuous curve is a guide to the eye. {(b) Inset: Log-log plot of the $C_c$ peaks for different system sizes ($N$). The grey dashed lines are fit to the points, with slope=$0.81, 0.74$ and $0.63$, for $E_{nn}=0.06,0.13$ and $0.17$, respectively.}}
\label{fig4}
\end{figure}

As the melting curve {crosses} the collapse line ({at $E_{nn}=0.2$ and $T=0.742$}), the {temperature induced melting} transition is gradually smeared with the order parameter smoothly going to zero. This gradual smearing may indicate the onset of an infinite-order transition. For finite-order phase transitions, the finite-size scaling (FSS) is a powerful tool to extrapolate the power-law behavior of the thermodynamic observables around the criticality. It is performed by scaling the distance to the critical point and the observable height as functions of the system size and the critical exponents. However, the use of FSS is only restricted to problems with one diverging correlation length. Therefore, as the melting curve approaches the collapse line, with the emergence of another diverging length scale, besides that for melting, a proper estimate of the melting temperature becomes difficult, which limits our use of the FSS little above $E_{nn}=0.2$. Beyond this value, we find it difficult to perform a good data collapse, apparently because the two arms of the $C_c$ curves about the central peak collapses for two different values of $T_m$, denoting a range of $T$ values for the melting transition. However, such observations could be the result of poor performance and therefore sampling in the collapsed region or {may} require the usage of extended finite-size scaling (EFSS) analysis \cite{santos1981}. Notice that, considering the system is placed at a temperature (say $T=0.741$) which is considered to be a bound phase in the good solvent, the system can be carried to a completely unbound state just by changing the solvent quality at a fixed temperature, and much before the collapse transition [Fig.~\ref{fig3}].

In a good solvent, the energetic gain over entropy leads to a bound phase, thereby, lowering the free energy. Now, in the presence of nn attraction, there is an added energetic gain. However, the nn interactions play a dual role {in modifying} the melting transition. While $\text{nn}_{\text{ds}}$ should help to maintain the bound phase, $\text{nn}_{\text{ss}}$ should support the unbound phase, thereby, stabilizing the unbound phase. To supplement the fact that the initial (i.e for small values of $E_{nn}$) decrease in $T_m$ is solely due to the $\text{nn}_{\text{ss}}$, we carried out separate simulations where we turned off the $\text{nn}_{\text{ss}}$ interactions. We found that the melting temperature only increases, even for small $E_{nn}$ values.  {The $\text{nn}_\text{ss}$ interactions are contributed by the broken bonds from the thermal bubbles and the Y-fork at the end.} Since, for small values of $E_{nn}$, {in the vicinity of melting point} \cite{comm2}, configurations with large bubbles that can contribute to $\text{nn}_{\text{ss}}$ should be rare, it is the mechanism by which the Y-fork opens up that ultimately determines the $\text{nn}_{\text{ss}}$. 
\begin{figure}
\centering
\hspace*{-7cm}(a) \\
\includegraphics[width=.9\linewidth]{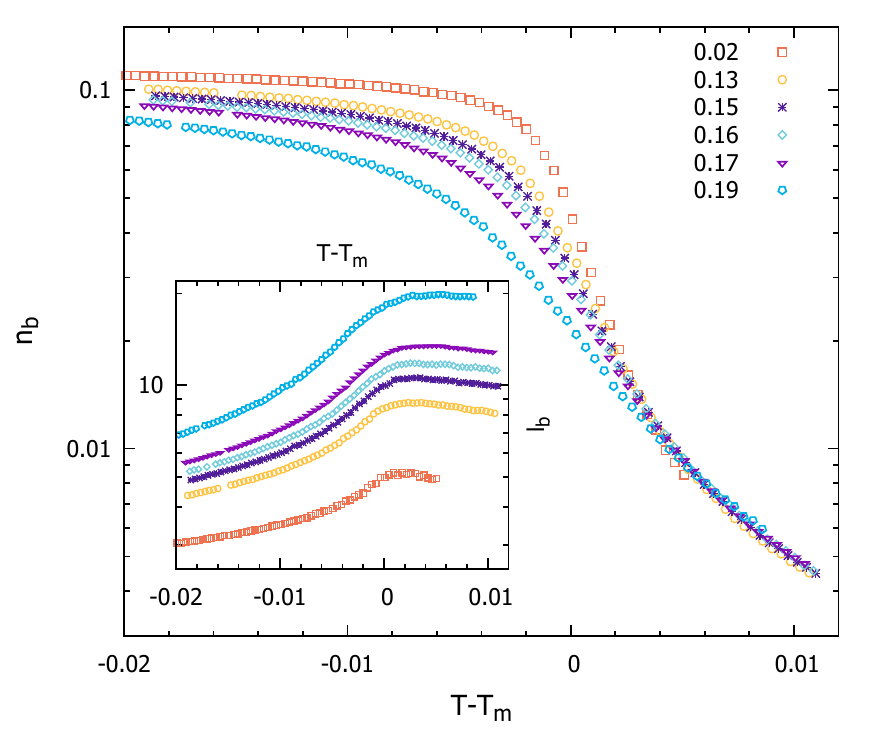}\\
\hspace*{-7cm}(b)\\
\includegraphics[width=.9\linewidth]{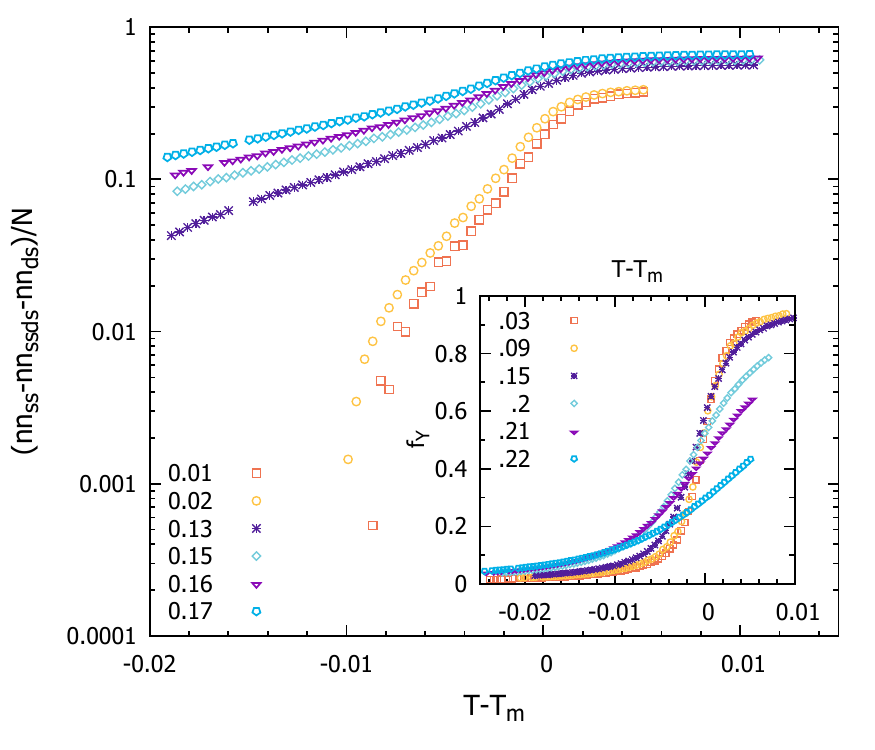}
\caption{(Color online) (a) Average number of bubbles per unit length ($n_b$) for chain length $N=2000$. (Inset) {Semi-log plot of} average bubble size ($\ell_b$) along the chain, excluding the Y-fork, for different values of $E_{nn}$ and chain length $N=2000$. {For $\ell_b$}, average is taken along the contour length and then over many such configurations. The curves are shifted {by} their corresponding $T_m$'s to make the melting point coincide with zero. (b) Semi-log plot of the difference between the nn contributions from the ss and ds segments per unit length ($N$). The negative part of the curves for $E_{nn}=0.01$ and $0.02$ are not shown. (Inset) Fraction of broken bonds per unit length in the Y-fork ($f_Y$), for different values of $E_{nn}$. 
}
\label{fig5}
\end{figure}
On the other hand, with increasing $E_{nn}$, the average number of bubbles per unit length $(n_b)$ decreases [Fig.~\ref{fig5}(a)], while the average bubble length ($\ell_b$) along the chain increases [Fig.~\ref{fig5}(a) inset], along with a delayed Y-fork ($f_Y$) opening around the melting transition [Fig.~\ref{fig5}(b) inset]. This picture is consistent with the melting of Gaussian strands where the majority of the broken bonds are in the bubbles pre-melting \cite{majumdar2020}. This delay in the Y-fork opening [Fig.~\ref{fig5}(b) inset], along with a decreasing number and increasing average size of bubbles, gives rise to a stabilizing effect for sufficiently large $E_{nn}$ values ($E_{nn} \gtrsim 0.15$) upto the collapse line [Fig.~\ref{fig3}] \cite{majumdar2020}. Further, deep in the collapsed phase ($E_{nn}\sim 1.0$) {and within the studied temperature window ($0.73<T<0.75$)}, there seems to be no transition at all \cite{mikhailenko2000}, {with the DNA in an unbound phase}.

{From the discussion in the preceding paragraph,} we expect, that the slope of the phase curve (i.e. $\frac{dE_{nn}}{dT}$) to be determined by a quantity that changes sign as we move along the 1-dimensional critical manifold [Fig.~\ref{fig3}]. One such quantity would be the difference between the {various components contributing to nn interactions,} i.e., $\text{nn}_{\text{ss}}$ and $\text{nn}_{\text{ds}}$ [Fig.~\ref{fig5}(b)]. It is positive for large values of $E_{nn}$ on the bound side of the corresponding $T_m$'s, and rapidly changes sign for smaller $E_{nn}$ values, as expected for a sharper transition. From a thermodynamic viewpoint, this is equivalent to the Clasius-Clapeyron equation, $\frac{dE_{nn}}{dT}=\frac{L}{T\Delta nn}$, where $L=T(S_{\text{ss}}-S_{\text{ds}})$ {is the latent heat,}  $S$ is the entropy, and $\Delta nn = (nn_{\text{ss}}-nn_{\text{ss-ds}}-nn_{\text{ds}})$. {Below the curve satisfying $E_{nn}/T = 0.269$}, polymers are in the swollen phase. Therefore, both $nn_{\text{ds}}$ and $nn_{\text{ss}} \rightarrow 0$ for $N\rightarrow \infty$. If there is latent heat i.e. $L\neq 0$, the slope would be $\infty$, i.e., the phase curve is a vertical line. If $L=0$, {as for continuous transitions}, then for finite $N$, $S_{\text{ss}}-S_{\text{ds}}\sim N^{a}>0$ with $a<1$, so that the entropy difference per unit length $\delta S/N \rightarrow 0$. So, we must have $\Delta nn \sim N^a$, for $N$ dependence to cancel, and if $\Delta nn<0$ on the phase boundary then $\frac{dE_{nn}}{dT}<0$. This inequality will change in the immediate vicinity of the critical points for larger $E_{nn}$ values, for the slope to become positive [Fig.~\ref{fig5}(b)].

For an explanation of the continuous variation of the crossover exponent $\phi$, we note that, pre-condensation, the two main length scales of the problem is given by the two correlation lengths related to the geometry of the polymeric strands $(\zeta_1)$ and another related to the mean diameter of the thermal bubbles $(\zeta_2)$ which is expected to be related to the average bubble length ($\ell_b$)  \cite{causo2000}. The critical exponents $\nu$ and $\nu_T$, control these length scales, {$\zeta_1$ and $\zeta_2$ respectively}, and using which we can write \cite{causo2000, santos1981},

\begin{equation}
\phi = \nu / \nu_T.
\end{equation} 

Now, with increasing $E_{nn}$, $\nu_T$ increases with an increase in the average bubble length $(\ell_b)$ along the chain [Fig.~\ref{fig5}(b) inset], while  $\nu = 1/3$ only as the melting curve crosses the collapse line, in the thermodynamic limit ($N\rightarrow \infty$), thereby leading to a gradual lowering in the value of $\phi$. In other words, the introduction of {effective attraction due to poor solvent, introduces long-range interaction among monomers which are far apart along the chain contour. This,} changes the way the correlations become infinite as $T\rightarrow T_m^{-}$, $\zeta_2\sim (T-T_m)^{-\nu_T}$, thus {denoting a change in} the exponent {$\nu_T$}\cite{baxter_book}.

Again, for continuous transitions, $\phi$ can be estimated from the bubble-size distribution (bsd) at the critical point, using the expression $P(\ell_b)\sim \ell_b^{-c}$, where $c=\phi+1$ is the bubble-size exponent (bse) \cite{carlon2002}. In Fig.~\ref{fig6} we have plotted the bsd at the critical points of the melting transition for $E_{nn}=0.03, 0.15, 0.17, 0.18, 0.19$ and $0.2$, which clearly shows that $\phi$ decreases with increasing $E_{nn}$. This, corroborates the fact that the continuous change in $\phi$ is, in fact a genuine change and not some artifact of the FSS originally used for estimation of $\phi$. However, we see differences in the value of the $\phi$ obtained from the two different methods. Interestingly, according to the estimation of $\phi$ from bse ($c$), there should be another tricritical point along the melting curve at $c=2$, {apparently where the melting curve changes slope.}

\begin{figure}[t]
\centering
\includegraphics[width=0.9\linewidth]{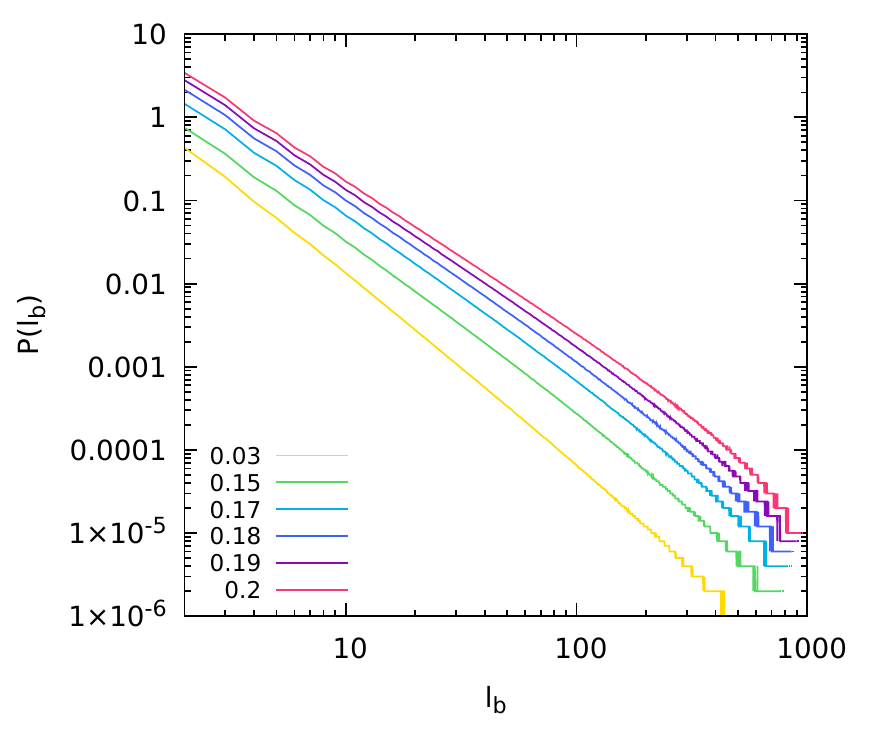}
\caption{Bubble-size $(\ell_b)$ distribution $(P(\ell_b))$ at the critical points for $E_{nn}=0.03, 0.15, 0.17, 0.18, 0.19$ and $0.2$. Data starting from $E_{nn}=0.15$ are shifted by factor of $2$ to $10$ in steps of $2$. Fitting of the data is performed in the range $\ell_b=20$ to $200$, which gives $c=2.3, 2.07, 1.99, 1.94, 1.89$ and $1.83$ for increasing $E_{nn}$ starting from $0.03$ respectively.}
\label{fig6}
\end{figure}

In the study of critical phenomena, a continuous variation of the exponents, often induced by a marginal interaction, in the {renormalization group} sense, is rarely observed \cite{khan2017, martins2018}. For the unperturbed case, the two universality classes for melting, using  PS and alike models, are for ideal random walk ($\phi=1/2$ and continuous melting) and self-avoiding walk ($\phi=1$ and first-order melting). The addition of solvent interaction acts as a relevant perturbation that brings the system out from its universality class as of the unperturbed case. However, whether the transition is non-universal or there is an underlying weak-universality, where the critical exponents vary, but the ratio of two critical exponents remains constant, is to be investigated further. Such a change in the critical exponents is not new and was observed previously upon adding the solvent effect for adsorption of polymers on surface \cite{martins2018}.

\section{Conclusion} 
\label{sec5}
In conclusion, we studied the possibility of various phases resulting from the interplay between the base-pairing and nn interaction due to poor solvent. It {is evident} that the melting transition is continuous, even with excluded volume interaction, for finite nn interaction {and much before the collapse transition}. As we decrease the solvent quality, initially, there is a decrease in the melting temperature, followed by {a subsequent} increase in the later stages as the melting curve approaches the collapse line. The slope of the melting curve is determined by a competition between the nearest-neighbor interactions from the double-stranded and single-stranded segments. While scaling holds in the coil region, universality does not, {leading to a continuously varying critical exponent} \cite{baxter_book}. We provide evidence from three different methods to support this fact; first, from the FSS; second, from the average size of bubbles; and third from the bubble-size distribution. Apparently, there is a gradual crossover between the two universality classes for DNA melting. Melting as a transition is suppressed in the deep-collapsed region with the breakdown of FSS. However, there could be a smooth crossover between the bound and unbound phases over a large temperature range. Apart from that, we also see a possibility of solvent-mediated melting. Importantly, our results are in qualitative agreement with the previous experimental results \cite{baldini1985, hammouda2009}.

 Including phase separation effects in the solvent could be one of the interesting extensions of this work worth {investigating}. {One can easily extend this model} to include different rigidity for the ss and ds segments \cite{majumdar2021}. Adding rigidity to the ds segments can change the results significantly since, the solvent will affect the rigid and flexible {components} differently \cite{bastolla1997}. {The} change in the effective excluded volume interaction should also change the bulk elastic response \cite{majumdar2020}. Note that in a real situation, the Y-fork can exist at both ends, due to which the slope of the melting curve might change. {Our results would help set the {\it in vitro} environmental conditions for the experiments.}

Finally, it is often seen that biological systems at all scales are often poised at criticality, which can be both statistical and dynamical. Maintaining itself at the criticality has its advantages, e.g., high susceptibility to external perturbations \cite{mora2011}. However, how this critical point is searched for and maintained by these biological systems is another puzzle since the critical region spans only around a small region (or a single point in the thermodynamic limit) of the parameter space. The possibility that DNA might poise itself near the critical point or if a diagonal line like in Fig.~\ref{fig2} is possible is something that needs to be investigated and how it can be advantageous for the system, given that DNA is such an important functional unit of the cell.

\section*{Data Availability}
Data and information about the simulations are available upon reasonable request to the author.

\section*{Conflict of Interest}
The author has no conflict of interest to declare that are relevant to the content of this article.

\section*{Acknowledgement}
DM is indebted to Somendra M Bhattacharjee for insightful discussions. Simulations were carried out at the \textsc{SAMKHYA} computing facility at IOPB.


\begin{thebibliography}{99}
\bibitem{comm1} In eukaryotes, however, the folding mechanism is rather sophisticated and requires the involvement of histomers.
\end{thebibliography}
\end{document}